# Cancer Risk Messages: Public Health and Economic Welfare

Ruth F. G. WILLIAMS [a,b,1], Ka C. CHAN [c], Christopher T. LENARD [a] and Terence M. MILLS [a]

[a] *Department of Mathematics and Statistics, La Trobe University Bendigo, VIC 3552, Australia*
[b] *The Graduate School of Education, The University of Melbourne, VIC 3052 Australia*
[c] *School of Business, Education, Law and Arts, University of Southern Queensland Springfield Central, QLD 4300, Australia*

**Abstract.** Statements for public health purposes such as "1 in 2 will get cancer by age 85" have appeared in public spaces. The meaning drawn from such statements affects economic welfare, not just public health. Both markets and government use risk information on all kinds of risks; useful information can, in turn, improve economic welfare, however inaccuracy can lower it. We adapt the contingency table approach so that a quoted risk is cross-classified with the states of nature. We show that bureaucratic objective functions regarding the accuracy of a reported cancer risk can then be stated.

## Introduction

Public health messages quoting the risk of cancer are commonly issued. For example, in Australia, "By the age of 85, the risk is estimated to increase to 1 in 2 for males and 1 in 3 for females" [1]. Cancer risk messages are also stated on cancer websites internationally.

The concept of cumulative risk underlies these "1 in 2" and "1 in 3" types of messages (regarding Australia) and assumes cancer is the only cause of death [2–5]. Since the reported risk of cancer in Australia has encompassed this implicit assumption, an over-statement of the risk is given by this message. Australia is not the only country overlooking this assumption in messages to the public about the quoted risk of cancer.

Cancer risk information protects public health. As well, risk of cancer messages affect resource allocation overall, and thus economic welfare. Some of the economic implications of risk messages are being overlooked in general, and the quality of the cancer risk information being quoted and reported also affects resource allocation. Herein an *a priori* reasoning exercise clarifies the approach to improving the quality of the risk information being quoted and reported on cancer. It should be noted that the focus here is on the objective content, or accuracy, of cancer risk messages provided to the public: personal responses to cancer risk messages, such as positive and negative motivations, fear, anxiety, fatalism etc, are a separate topic.

---

[1] Corresponding Author: Dr Ruth Williams; E-mail: ruth.williams@latrobe.edu.au; williams.r1@unimelb.edu.au

An historic illustration is useful to show the implication for economic welfare of a policy stance towards a health risk. The example is the epidemiological evidence about the risk level of the obstetric X-ray. The relevant evidence which was discovered by Alice Stewart in the 1950s involving a theoretical error at that time that had developed amongst scientists who were misconceiving threshold levels. Contrary to theory current at that time, the tipping point for obstetric X-ray is **not** a value **greater than** zero: the tipping point value **is** zero. The effect overall on economic welfare of miscalculating the public health risk of obstetric X-rays was not trivial. Foremost, the incidence of childhood leukaemia heightened. Implicitly there existed a severe under-allocation of resources to the public message about the risk of exposing pregnant women to X-rays. Zero resources were spent on that message whilst far too many resources were directed to obstetric X-ray.

This historic illustration about a risk estimate that was objectively incorrect in the case of obstetric X-rays tells of incorrect information on the false negative rate, and affected the message taken from the risk signal issued**:** that incorrect information was not apparent. The rate was rising, and unnoticed; and, in turn, scarce pregnancy resources were increasingly misallocated through time. Incidentally, the appropriate public message for the (then) practice of obstetric X-ray, *viz.* that practice ought to be abandoned, formed only after three decades [6].

On economic grounds, the judicious management of cancer resources encompasses the notion of useful public knowledge about the risk of cancer. This idea is a new notion in current times. The relevant economic literature on health is also not new. Regarding the former, daily life in many countries proceeds on a base of useful literacy about everyday risks. This literacy exists because risk information has economic value to populations. There are many examples of informing the public of risks, such as: macroeconomic uncertainty; meteorological risk (e.g. cyclones and tornadoes) and planetary climate; natural hazards such as bushfires and seismic disturbance; financial safety and consumer protection policy; and all the umbrellas for public safety such as public health, workplace safety and road safety. The economic decision-making over health involves many types of risks, and the economic decisions that arise from those risks, such as participating in cancer screening services, or not doing so; fitness, or not; exercising; nutritious food, or not; smoking, or not; excessive consumption of alcohol, or not; being SunSmart, or not.  The decisions over all such dichotomies involve opportunity costs; and there is a spectrum of stances in between the dichotomies: there is an opportunity cost with every decision. Stark errors and unrealistic assumptions in the risk of cancer messages to the public not only affect the health status at the population level but overall resource allocation, with flow-on effects for population mortality and life expectancy.

As mentioned in the previous paragraph, the economic literature on health the relevant theoretical development is no novelty in health economics. Those developments hearken to Grossman, Becker, Muth, and Pollak and Wachter [7–10]. The theoretical contributions about the link between information and economic efficiency began with Arrow, and Greenwald and Stiglitz [11–12]. Space limitations constrain discussion either of these developments or of the empirical studies that followed. We also place aside other conceptual developments about risk in the literature, including risk aversion and expected utility, as well as the developments reported in the psychology literatures about the limitations on human rationality as well as on processing the meaning of risk messages.

## 1. The economic dimensions of a reported cancer risk

Our study is a specific application to cancer risk message of developments in information economics that were informed by health economics and epidemiology. Economic links were measured between the informational content of alternative diagnostic tests of the gastrointestinal tract, specifically the predictive value of the existing and new diagnostic tests, and health expenditures [13]. Doessel's study draws on concepts on informational accuracy already in epidemiology due to Yerushalmy [13, 14], concepts that are now standard fare in epidemiology courses [15]. The original concepts imply that there is a cross-classification behind determining sensitivity and specificity information about diagnostic procedures [14]. Vecchio's subsequent analysis is a further development [16]. Thus, the conceptual and empirical evidence for the economic implications of the epidemiological developments exists [13–15].

The context of these developments is the act of comparing the informational value of **alternative** diagnostic tests and there are clinical and economic implications of that information. For a clinician, the dilemma of the information provided by a diagnostic test result is to know whether, or not, the patient actually has the disease having been tested for, given the test result. The concepts for the clinical situation are as follows. There is the positive predictive value (*PPV*), which is the probability that subjects with a positive test truly have the disease, and there is negative predictive value (*NPV*), which is the probability that subjects with a negative test truly do not have the disease. These developments arose in the context of diagnostic test information and even though a public health message is not a diagnostic test, the relevant economic concepts for cancer risk messages are also in these findings. The information conveyed publicly by risk of cancer messages is a self-diagnosed personal risk of cancer. The predictive accuracy is thus subject to considerable uncertainty.

For our purposes an individual draws a datum on one's cancer risk from a reported population risk of cancer. For our precise focus, let us focus only on a single specific signal being received by an individual from a quoted cancer risk in a public message. We extend Doessel's approach in order to inject economic insights to public health messages. To explain, it is necessary to simplify the information that a person can draw from a public health message about the risk of cancer.

Let us assume that two alternative, or dichotomous, conclusions are drawn: either a person concludes that "I have a high risk of cancer" or "I have a low risk of cancer". As with the cross-classification approach to diagnostic tests, it will be necessary to bear in mind as well that there are two actual states of nature: either the person gets cancer by age 85 or the person does not get cancer by age 85.

Let us next consider the case of an inaccurately reported risk. In any risk information conveyed to the public, inaccuracy may be experienced by the individual, from three sources at least: the risk reported to the population, the error in the risk information at the population level, and the error in the risk information at the individual level. The specific focus in the present study is on the first source only, inaccuracy in the quoted risk.

We explicate the specific economic effect of this inaccuracy by an application of Doessel's contingency table (or confusion matrix) [13]. Refer to Table 1. This table indicates the implications of inaccuracy in quoting cancer risk: an economic focus has the opportunity costs in mind of alternative actions towards cancer prevention, which are based on one's risk information. This table is similar to the confusion matrices used in conceiving of the information from diagnostic testing, *viz.* there are the actual states and there is the informational content of a risk message. The cross-classification presented in Table 1 as a confusion matrix depicts the four alternative outcomes from the risk of cancer message, therein incorporating how to conceive of the "signal error", and alternative allocations of resources are implied. These classes are shown in the column and row headings. There are consequences or outcomes that are four fold: true positive (*TP*), false positive (*FP*), true negative (*TN*) and false negative (*FN*). Note that these outcomes of the individual information about the risk of cancer derived from the quoted population risk are measured in probabilistic terms: the predictive value of a positive (high cancer risk) message and the predictive value of a negative (low cancer risk) message.

**Table 1.** Outcomes from a risk of cancer message.

| | | *Information result of message* | | |
| --- | --- | --- | --- | --- |
| | | The people perceiving their cancer risk is high | The people perceiving their cancer risk is low | Row totals |
| *Actual cancer states* | The people who get cancer by age 85 | *TP* | *FN* | *TP* + *FN* = People who get cancer by age 85 |
| | The people who do not get cancer by age 85 | *FP* | *TN* | *FP* + *TN* = People who do not get cancer by age 85 |
| | Column totals | A high risk message being received[§] = *TP* + *FP* | A low risk message being received[§] = *FN* + *TN* | Total population at risk = *TP* + *FN* + *FP* + *TN* |

*Notes*:
*TP* is true positives, i.e. persons who get cancer by age 85 for whom the risk of cancer message means high cancer risk.
*FN* is false negatives, i.e. persons who get cancer by age 85 for whom the risk of cancer message means low cancer risk.
*FP* is false positives, i.e. persons who do not get cancer by age 85 for whom the risk of cancer message means high cancer risk.
*TN* is true negatives, i.e. persons who do not get cancer by age 85 for whom the risk of cancer message means low cancer risk.
[§] Risk aversion behaviour etc. is placed aside for now.

Despite the conceptual similarities with diagnostic test information, there is not any tidy datum about one's personal cancer risk that is available from a public health message with current technology; and thus a person is encouraged to ameliorate genetic and age tendencies to cancer onset by responding to a reported risk. Although *FN* and *FP* error are conceived of conventionally, note that the *TP* class for a cancer risk message incorporates those who, though believing they have high cancer risk and acting accordingly to lessen their risk, are yet unsuccessful in any preventive efforts that postpone onset; and there are also others whose health actions serve only to exacerbate their cancer risk. Also, the *TN* class is merely a situation of those who do not seek to prevent cancer and do not die from cancer – death by age 85 happens nonetheless, from another cause.

## 2. The two parts of the policy objective for cancer risk information

Table 1 also indicates the context of rational policy over cancer risk information. It is possible to incorporate economic welfare into the decision-making that arises from cancer risk information. Rational policy over risk of cancer messages considers the opportunity costs of policy alternatives. At first glance, placing a focus solely on the *TP* cell in Table 1 could seem reasonable. A single focus for cancer policy, such as a focus of the TP rate, appears to define the bounds of a risk message that informs people accurately of their cancer risk. However, a focus just on the *TP* rate does not avert losses to economic welfare from inaccurate information about the risk of cancer message: resource misallocation is likely from an inaccurate risk value. The confusion matrix implies that an objective that is set to lower the FN rate [which is $FN/(FN + TP)$] from cancer risk messages promote both public health and also raise economic welfare. The future (actual) cancer state of the population is not known presently; resources are least misallocated from economic decisions made by individuals receiving a cancer risk signal when the inaccuracy of the message is the lowest possible within a given budget constraint.

Let us express this notion in terms of a message having predictive value, and recall

$$NPV = \frac{TN}{FN+TN} \tag{1a}$$

and it can be stated also that

$$(1 - NPV) = \frac{FN}{FN+TN} \tag{1b}$$

It follows that the bureaucratic objective function for a cancer policy that incorporates policy on messages about the risk of cancer is:

$$Min. (1 - NPV) \text{ subject to the budget constraint.} \tag{2}$$

This objective encompasses averting the risk of cancer as well as averting losses to economic welfare by a policy focus that incorporates appropriately minimising inaccuracy on information on cancer risk.

For rational public policy, a further public health objective is implied in the objective of minimising $(1 - NPV)$, within a given public health budget. The public health policy maker cannot focus on the cancer risk message only, as if there are no other causes of death. A two-part policy objective involves lowering not only the accurately reported TP rate but also minimising (1-NPV) subject to the budget constraint. The two parts of the objective ensure jointly that the policy maker avoids heightening the risk of mortality from other causes. This is a possible implication of a cancer risk message that has low accuracy that in turn misallocated resources.

The two parts are mutually exclusive, though inverse, within the existing mortality rate. Ignoring some cells in the contingency table is hazardous to economic welfare, which can cause flow-on effects to public health anyway. Overall economic efficiency can be improved from accurate cancer risk messages by reducing both erroneous false positive information **and** erroneous false negative information.

## 3. Conclusions

We have examined whether, or not, accurate messages about cancer risk can improve not only public health, but economic welfare also. This exposition concludes with an affirmative response *a priori*: there are economic implications of cancer risk messages.

The implications under consideration arise in the private and the public sectors of the economy as risk messages involve informational balances affecting economic welfare. There is both a false negative rate and a false positive rate to be considered in the risk being quoted. Private decisions in response to a reported risk affect economic welfare for everyone as private decision-making affects resource allocation overall. In the public sphere, policy makers either under-respond or they can over-respond in any policy stance adopted towards public health information reporting the risk of cancer. Following Niskanen, it is vital for governments to restrain bureaucrats from budget maximisation in public health, ensuring instead that public health budgets pursue social efficiency [17]. The argument herein is somewhat unsurprising: the quoted risk of cancer obviously is relevant to individual economic responses to cancer risk. However policy makers ought to be governed by the appropriate objective/s. More work needs to be done but with some initial results, we can demonstrate the appropriate approach to risk of cancer estimation. The model can be developed further to answer many useful questions with appropriate assumptions.